\documentclass[twocolumn]{aa}
\usepackage{graphicx}
\usepackage{txfonts}
\begin{document}
   \title{The VIMOS VLT Deep Survey}


   \subtitle{Public release of 1599 redshifts to $I_{AB}\leq24$
across the Chandra Deep Field South
\thanks{The data presented in this paper has been 
         obtained with the European Southern Observatory Very Large
         Telescope, Paranal, Chile }
}

   \author{
O. Le F\`evre \inst{1}, 
G. Vettolani \inst{2},
S. Paltani \inst{1},
L. Tresse  \inst{1},
G. Zamorani  \inst{10}, 
V. Le Brun \inst{1},
C. Moreau \inst{1},
C. Adami \inst{1}, 
M. Arnaboldi \inst{7}, 
S. Arnouts \inst{1}
S. Bardelli \inst{10},
M. Bolzonella \inst{11},
M. Bondi \inst{2}, 
A. Bongiorno \inst{11},
D. Bottini \inst{3}, 
G. Busarello \inst{7}, 
A. Cappi \inst{10}, 
P. Ciliegi\inst{10} , 
T. Contini \inst{4},
S. Charlot \inst{5}, 
S. Foucaud \inst{3},  
P. Franzetti \inst{3},
B. Garilli \inst{3}, 
I. Gavignaud \inst{4}, 
L. Guzzo \inst{6}, 
O. Ilbert \inst{1}, 
A. Iovino \inst{6}, 
D. Maccagni \inst{3}, 
D. Mancini \inst{7},
B. Marano \inst{11}, 
C. Marinoni \inst{1},
H.J. McCracken \inst{8}, 
G. Mathez \inst{4}, 
A. Mazure \inst{1},
Y. Mellier \inst{8}, 
B. Meneux \inst{1},
P. Merluzzi \inst{7},
C. Moreau \inst{1}, 
R. Pell\`o \inst{4}, 
J.P. Picat \inst{4},  
A. Pollo \inst{6}, 
L. Pozzetti \inst{10},
M. Radovich \inst{7}, 
V. Ripepi \inst{7},
D. Rizzo \inst{4}, 
R. Scaramella \inst{2}, 
M. Scodeggio \inst{3}, 
A. Zanichelli  \inst{2}, 
E. Zucca  \inst{10}
          }

   \offprints{O. Le F\`evre}

   \institute{
Laboratoire d'Astrophysique de Marseille, UMR 6110 CNRS-Universit\'e
de Provence, Traverse 
    du Siphon-Les trois Lucs, 13012 Marseille, France\\
              email: olivier.lefevre@oamp.fr
         \and
Istituto di Radio-Astronomia - CNR, Bologna, Italy
\and
IASF - INAF, Milano, Italy
\and
Laboratoire d'Astrophysique - Observatoire Midi-Pyr\'en\'ees, Toulouse, France
\and
Max Planck Institut fur Astrophysik, 85741 Garching, Germany
\and
Osservatorio Astronomico di Brera, via Brera, Milan, Italy
\and
Osservatorio Astronomico di Capodimonte, via Moiariello 16, 80131 Napoli, Italy
\and
Institut d'Astrophysique de Paris, UMR 7095, 98 bis Bvd Arago, 75014 Paris, France
\and
Observatoire de Paris, LERMA, UMR 8112, 61 Av. de l'Observatoire, 75014 Paris, France
\and
Osservatorio Astronomico di Bologna, via Ranzani 1, 40127 Bologna, Italy
\and 
Universit\`a di Bologna, Departimento di Astronomia, via Ranzani 1, 40127 Bologna, Italy
             }

   \date{Received ..., 2004; accepted ..., 2004}

   \abstract{This paper presents the VIMOS VLT Deep Survey
around the Chandra Deep Field South (CDFS). We have measured
1599 new redshifts with VIMOS on the European Observatory
Very Large Telescope - UT3,
in an area $21 \times 21.6$ arcmin$^2$, including 784
redshifts in the Hubble Space Telescope - Advanced Camera
for Surveys GOODS area. 30\% of all objects with $I_{AB}=24$
have been observed independently of magnitude, indicating
that the sample is purely
magnitude limited. We have reached
an unprecedented  completeness level of 88\% in terms of 
the ratio of secure measurements vs. observed objects,
while 95\% of all objects have a redshift measurement. 
A total of 1452 galaxies, 139 stars, 8 QSOs have a 
redshift identification, 141 of these being unsecure 
measurements. The redshift distribution down to $I_{AB}=24$
is peaked at a median redshift z=0.73, with a significant 
high redshift tail extending up to $\sim4$. Several high
density peaks in the distribution of galaxies are
identified. In particular, the strong peak at z=0.735 
contains more than 130 galaxies in a velocity range
$\pm2000$km/s distributed all across the transverse
$\sim$20 $h^{-1}$ Mpc of the survey. We are releasing all redshifts
to the community, along with the cross identification
with HST-ACS GOODS sources on the CENCOS database environment
http://cencosw.oamp.fr.
   }

\authorrunning{Le F\`evre, O., Vettolani, G., et al.}
\titlerunning{VVDS: public release of redshifts in CDFS}

   \maketitle
%

\section{Introduction}

Understanding the major steps in the evolution of galaxies
still remains a major challenge to modern astrophysics. While  
the general theoretical framework of the  hierarchical growth of 
structures in the universe including the build up of
galaxies is well in place (e.g. \cite{peacock04}), at high redshifts 
this picture remains largely unconstrained by
observations. The detailed properties of the main population of galaxies 
from large samples  representative of the universe at various 
epochs remain to be established
across most of the life of the universe
beyond the large local volumes explored by the 2dFGRS (\cite{colless}
and the SDSS (\cite{SDSS}), and expanding from smaller exploratory
surveys (\cite{lilly95}, \cite{lefevre95}, \cite{steidel03}, 
\cite{cimatti03}). 

The VIMOS VLT Deep Survey (VVDS) is a deep redshift survey
aimed at studying the evolution of galaxies, large scale structures and
AGNs over more than 90\% of the current age of the universe. The unique
feature of the VVDS is the simple magnitude selection applied 
to define a complete magnitude limited sample of distant galaxies, with
a goal of more than 100000 objects observed in
multi-object spectroscopy. The VVDS rests on the
observations of 5 different fields to smooth out the effects of 
cosmic variance when building the statistical properties of the
galaxy population (\cite{lefevre04}).

This paper presents the redshift survey observations of 1599 
objects with $I_{AB}\leq24$ 
conducted by the VVDS team around the Chandra Deep Field South,
including the HST-GOODS area (\cite{goods}). 
The observations have been carried out
with the VIsible Multi-Object Spectrograph (VIMOS) on the 8.2m 
Melipal telescope of the European Southern Observatory 
Very Large Telescope. We are describing the processing
steps and redshift measurements, and the associated quality
control we have applied to these data. The content of the final 
catalog is detailed, as well as the cross identification
with the Hubble Space Telescope Advanced Camera for Surveys GOODS 
images, and we present 
the main entries available from our interactive database.
The main properties of the sample are briefly
presented, including the redshift distribution of the sample.



\section{Observations}

\subsection{Multi-Object Spectroscopy with VIMOS}

Spectroscopic observations have been conducted with 
VIMOS on the VLT-UT3 Melipal (\cite{lefevre03}).
The low resolution red grism LRRED has been used with slits
of 1 arcsec width. The spectral resolution in this mode
is 34\AA at 7500\AA or $R\sim220$. The spectral length
has been limited by the red bandpass filter to $5500-9500$ \AA.
Slits placed on objects have a typical length $\sim10$ arcsec each.

\subsection{VIMOS pointings}

A complete VIMOS pointing is a combination of observations
with the 4 quadrants of the instrument, each separated by a 
cross about 2 arcminutes wide. With the above setup  and the
projected sky density of objects down to $I_{AB}=24$,
one VIMOS pointing allows to observe $\sim500-575$ targets
in one single observation (\cite{lefevre03}).

\begin{figure*}
   \centering
\includegraphics[width=9cm]{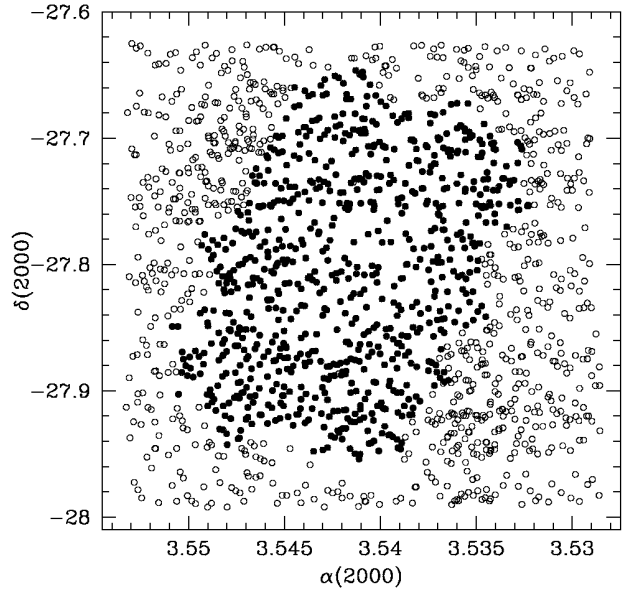}
      \caption{Objects observed with VIMOS-VLT around the
Chandra Deep Field South. Black circles are objects
in the HST-ACS GOODS area. 
              }
         \label{pointings}
   \end{figure*}

We have set a total of 5 pointings around the 
Chandra Deep Field South, the positions are listed
in Table \ref{vpoint}. Together, they cover a total area
of $\sim 453 \arcmin^2$ including the complete HST-GOODS survey 
field (\cite{goods}).
The layout of observed galaxies is presented in
Figure \ref{pointings}. 

   \begin{table*}
\centering
      \caption[]{Observed VIMOS pointings.}
         \label{vpoint}
     $$ 
         \begin{array}{lcccccc}
            \hline
            \noalign{\smallskip}
            Pointing     & \alpha_{2000} & \delta_{2000} &  Date & Number & T_{exp}   \\
            VVDS          & (2000)  & (2000) & Observed & of slits & minutes \\
            \noalign{\smallskip}
            \hline
            \noalign{\smallskip}
            CDFS001  & 03h32m28.0s & -27\degr48\arcmin30\arcsec & 31-Oct-02 & 447 & 10x27 \\
                     &     &                       & 1,2,4-Nov-02 & & \\
            CDFS002  & 03h32m37.04s & -27\degr50\arcmin30\arcsec & 5,6-Nov-02 & 331^{\mathrm{a}} & 8x27  \\
            CDFS003  & 03h32m18.95s & -27\degr50\arcmin30\arcsec & 9,10,11,12-Nov-02 & 447 & 9x27  \\
            CDFS004  & 03h32m37.04s & -27\degr46\arcmin30\arcsec & 27,28,29-Nov-02 & 436 & 12x27\\
                     &     &                       & 1,2-Dec-02 & &  \\
            CDFS005  & 03h32m18.95s & -27\degr46\arcmin30\arcsec & 2,4,5,6-Dec-02 & 448 & 10x27  \\
            \noalign{\smallskip}
            \hline
         \end{array}
     $$ 
	 \begin{list}{}{}
	 \item[$^{\mathrm{a}}$] Quadrant 2 not observed
	 \end{list}
   \end{table*}

\subsection{Mask preparation}

The preparation of slit masks for VIMOS observations has been
done using the photometric catalog produced by the 
ESO Imaging Survey (EIS, \cite{EIS}), and short images taken with VIMOS. 
The VIMOS images are used to produce a catalog of 
source positions in the VIMOS instrument coordinate system, which 
are then cross-correlated with the EIS catalog to compute the
transformation matrix from the EIS catalog astrometric 
system to the VIMOS focal plane where slit masks are located.
The VMMPS code was then run on the EIS catalog for all sources
brighter than $I_{AB}=24$ to optimize the number and positions
of slits for each of the 4 masks per pointing. Masks
have been cut by ESO Paranal Staff using the Mask Manufacturing
Unit (\cite{conti}).

\subsection{VIMOS observations}

Observations with VIMOS have been obtained between October 31
and December 6, 2002. Observing conditions were 
photometric with an image quality between 0.6 and 1.2 arcsec FWHM.
We have moved the telescope, hence the objects along the slits, in a sequence 
of 5 positions with offsets -0.7, -0.3, 0, +0.3, +0.7 arcsec from the
reference pointing position. This is necessary to compute  
the fringing pattern produced above 8300\AA ~by the thinned EEV CCDs 
used in VIMOS, and remove it during processing.

Wavelength calibrations have been obtained during the 
day, observing Helium and Argon arc lamps through the observed 
masks. The spectrophotometric standard star LTT3218 has been used
to derive the absolute flux calibration.

\section{Data Processing}

Data processing has been conducted under the VIPGI environment
developed by our team (\cite{scodeggio}). VIPGI has been
used to organize the multiple files and process all data
from the raw 2D images and calibration to the production
of sky subtracted, wavelength and flux calibrated 1D spectra.
Because of instrument flexures not yet minimized at the time 
of these observations, the fringing pattern has been occasionally
hard to remove. The quality control performed on these steps 
is described in \cite{lefevre04}. The wavelength accuracy is 
better than $\sim1$\AA ~rms all over the wavelength range,
and the spectrophotometry is accurate to about 10\%.

\section{Redshift Measurements}

Measuring redshifts for a complete magnitude limited sample
down to $I_{AB}=24$ had never been attempted before our
observations. The challenge is to measure redshifts
within a possible range $0 \leq z \leq \sim 5$, without any
a priori indication to preserve the complete magnitude 
selected sample approach. The approach we have followed
involves an iterative build up of galaxy templates 
as observed with VIMOS, coupled to the powerful redshift measuring
machine KBRED (\cite{scaramella}), based on cross correlation
and principal component analysis methods. This approach has
been applied and tuned on the more than 20000 spectra obtained
for the VVDS in the fall of 2002, and remained until recently
very manpower intensive. On this critical step,  we have
enforced a very strict quality control.

Each spectrum has been measured independently
by 2 VVDS team members, and then compared. A final check
has been done by a third team member prior to 
release into the database.
Each spectrum is assigned a redshift, and a flag indicating
the reliability level of the measurement, as defined in
\cite{lefevre95} and described in Section 6. 
Flags 2,3,4 are the most secure, 
flag 1 is an indicative measurement based on 
continuum and few supporting features,
and flag 0
indicates a measurement failure with no features identified. 
Flag 9 indicates
that there is only one secure emission line tentatively
assigned to the listed redshift (e.g. [OII]3727\AA ~or
H$\alpha$).

\subsection{Redshift accuracy}

\begin{figure*}
   \centering
      \includegraphics[width=9cm]{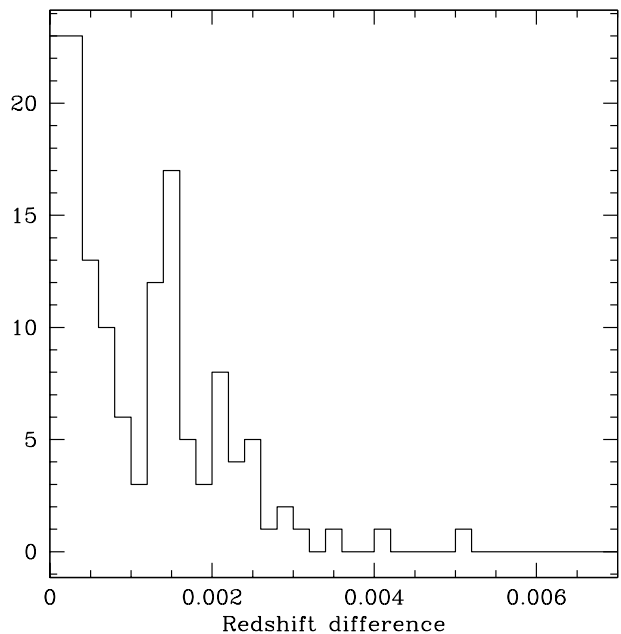}
      \caption{Redshift difference between objects observed
twice or more in independent VIMOS observations. The 
distribution has a velocity dispersion $\sigma_z=0.0012$ or $360$km/s. 
              }
         \label{diffz}
   \end{figure*}

The redshift accuracy can be estimated from a sample
of 160 galaxies which have been observed twice with
VIMOS within the 5 CDFS pointings. 
These galaxies have been included in two 
or more different mask sets, and observed 
independently at different times. The distribution
of measured redshift differences is presented in
Figure \ref{diffz}. The redshift dispersion of the distribution is
$\sigma=0.0012$, or 360km/s.  

We have compared our redshift measurements with the 
measurements of the K20 survey (\cite{cimatti03})
and VLT-FORS2 measurements conducted by the ESO-GOODS team
(www.eso.org/science/goods). A total of 70 objects
have been observed both by the VVDS and the K20.
For the 63 VVDS objects with flags $\geq 1$,
the redshifts agree for $\sim87$\% of the sample
with a $\Delta z=-0.0004$ and $\sigma_z=0.0017$,
with the main disagreement concentrated on flag 1,
as expected. The comparison with the FORS2 data
yields 46 objects in common, of which 42 have
a VVDS flag $\geq1$. The VVDS vs. FORS2 redshift
agree for 66\% of the sample with $\Delta z=-0.0006$ 
and $\sigma_z=0.0013$.
The detailed comparison of VVDS redshift measurements with
the measurements from these other teams will be presented in
\cite{lefevre04}.

\subsection{Completeness}

We have defined the completeness of the measurements as the 
ratio between the actual redshifts measurements and the
observed spectra. We have removed from the list of 
observed spectra those which have a clearly identified
instrumental or data processing problem affecting the 
measurement, like e.g. the slit is behind the guide probe of 
the VLT-UT3, or the data processing with VIPGI failed to
properly detect the object because of strong
residual features from the sky / fringing corrections.
This will be described in details in \cite{lefevre04}.

\begin{figure*}
   \centering
      \includegraphics[width=9cm]{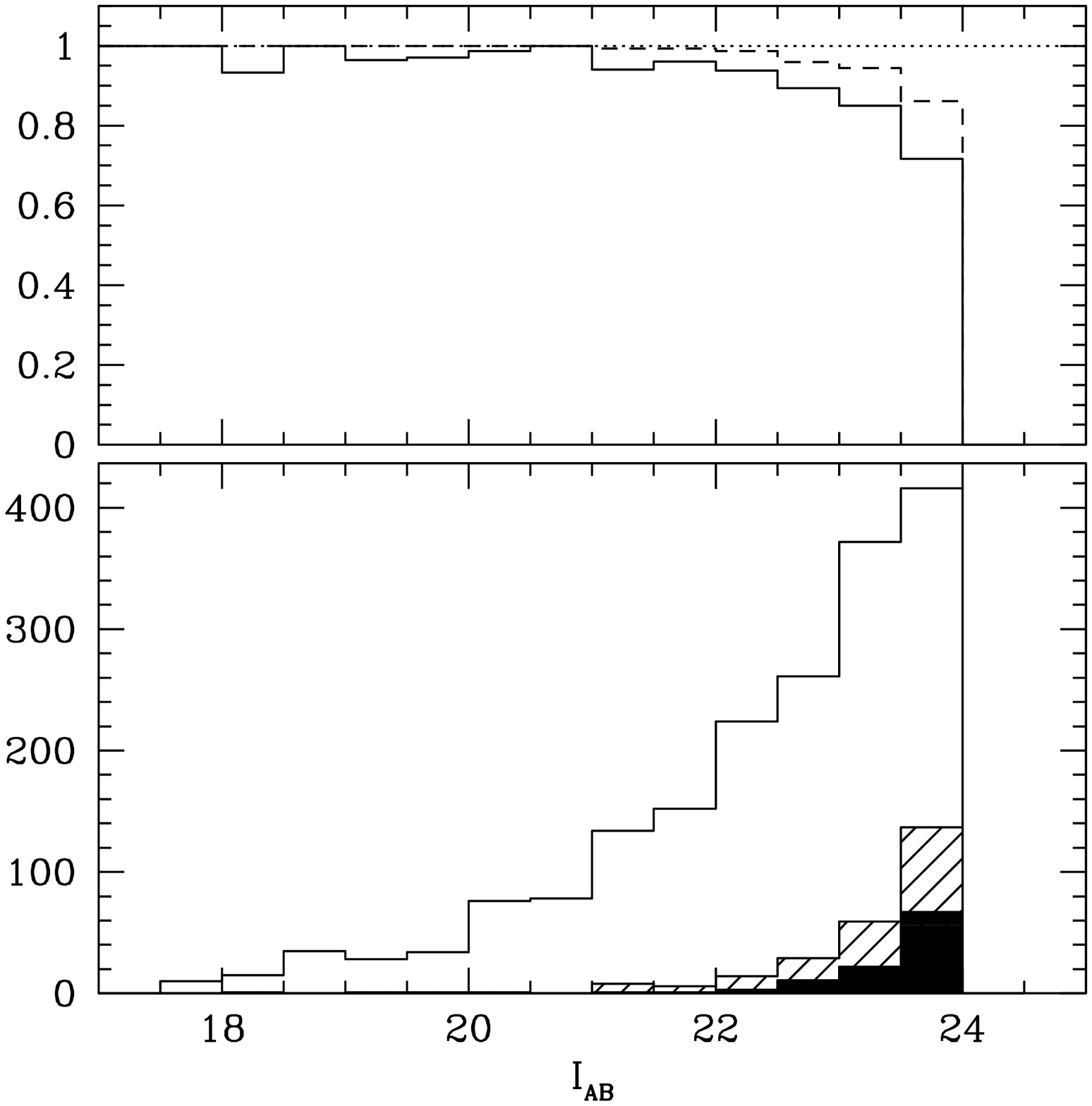}
      \caption{Completeness of the  $I_{AB} \leq 24$ sample. 
(Bottom panel) The magnitude distribution of galaxies
with secure redshift measurements (flags 2,3,4,9; open
histogram), uncertain and failed measurements (flags
0 and 1, dashed histogram), and failed measurements
alone (flags 0, filled histogram).
(Top panel) the
ratio of secure redshift measurements
(flags 2,3,4,9 continuous
line histogram), and of all measurements (flags 1,2,3,4,9
dashed line histogram). The overall redshift measurement
completeness is 88\% (flags 2,3,4,9) and redshifts
are measured for 95\% of the sample.
              }
         \label{completm}
   \end{figure*}

We are presenting in  Figure \ref{completm} the 
completeness vs. $I_{AB}$ magnitude of all objects
(galaxies, stars, QSOs) with flags 2,3,4,9 and 12,13,14,19
(the most secure) and the completeness level adding the
flags 1 and 11 to the above. 

The overall redshift measurement completeness reaches 88\% 
excluding flags 0 and 1; it reaches 95\% including objects
with flags 1 and excluding objects with flags 0.

\section{Cross identification with HST-ACS GOODS sources}

The cross identification of the objects observed from
the EIS catalog and the objects detected by HST-ACS
has been performed running a cross correlation of our
target list with the list of objects published in the 
version r1.0, December 22 2003, of the GOODS survey
multi-band catalog. The relative astrometry of the
EIS vs. GOODS catalogs has been found to be extremely good,
to within 0.1 arcsecond over our survey field. A search
circle with radius 0.3 arcsecond has been used to 
search for HST sources corresponding to the ground
based sources. This produced a matched
list with only one to one identifications, 
and no double identifications.

\section{Basic properties of the sample}

\subsection{Magnitude distribution}

\begin{figure*}
   \centering
      \includegraphics[width=9cm]{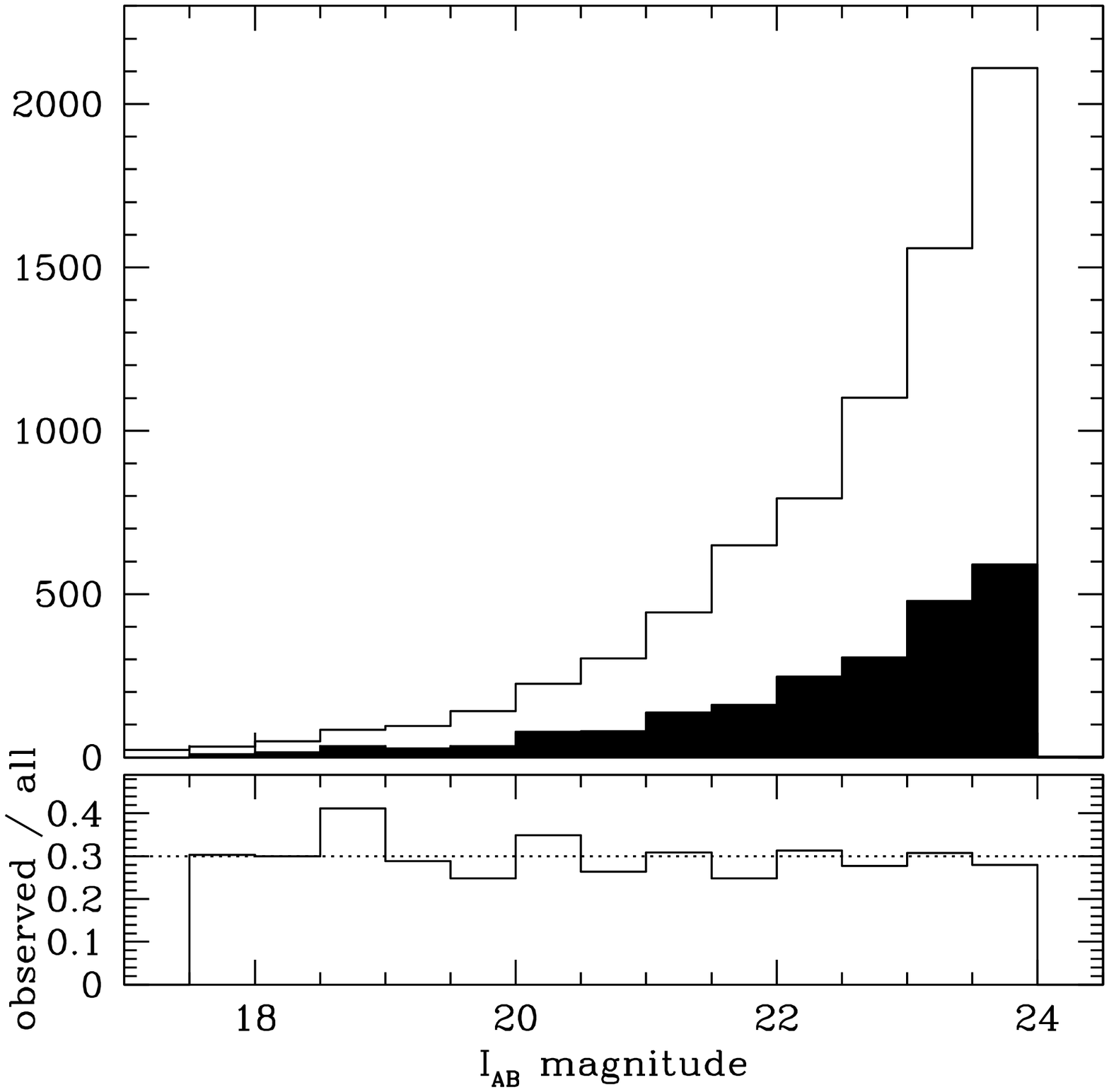}
      \caption{(Top panel) $I_{AB}$ magnitude distribution of objects
observed with VIMOS in the Chandra Deep Field 
South (shaded histogram), compared to the distribution
of all objects in the area in the EIS photometric
catalog (open histogram). (Bottom panel) The ratio of observed vs.
all objects is 0.3.
              }
         \label{magdist}
   \end{figure*}

The magnitude distribution of objects observed in
our survey is shown in Figure \ref{magdist}
compared to the distribution of all objects
in the EIS photometric catalog over the 
same area. 
We have observed 30\% of all objects with 
$I_{AB} \leq 24$ in the
area covered by the survey, independently
of magnitude, as shown in
Figure \ref{magdist}.
This demonstrates that there is no
magnitude - dependent bias in our object selection and  
that our sample is purely magnitude selected
even after the complex target selection in
the making of the VIMOS multi-slit masks.

\subsection{Redshift Distribution}

\begin{figure*}
   \centering
      \includegraphics[width=18cm]{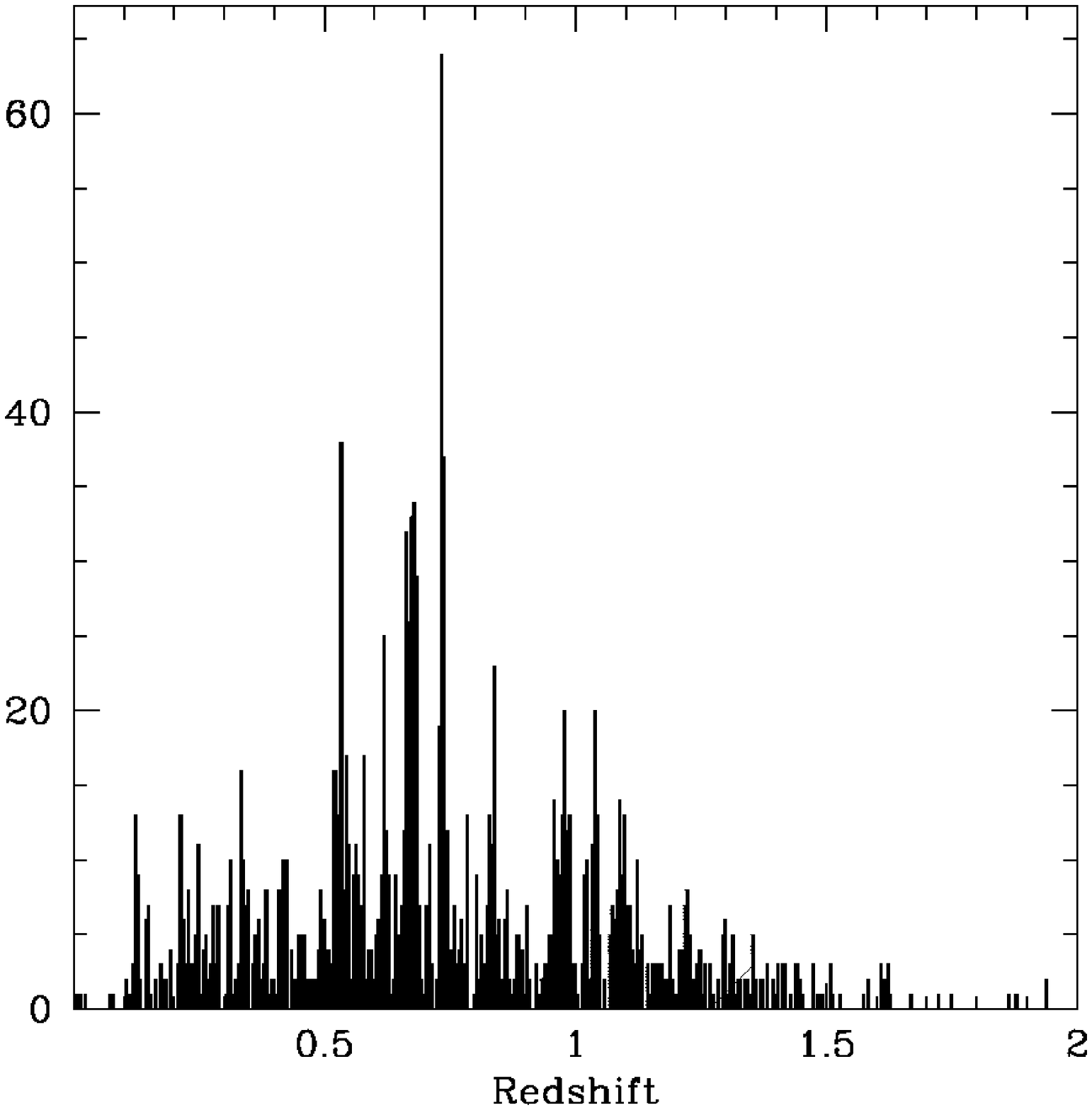}
      \caption{Redshift distribution of galaxies
               with $I_{AB}\leq24$ observed with
               VIMOS in
the Chandra Deep Field South. The redshift bin 
is dz=0.005. There 
are 42 objects identified with $z\geq2$ not included
in this figure.
              }
         \label{zdist}
   \end{figure*}

The redshift distribution of the full sample of 
galaxies and QSOs
is presented in Figure \ref{zdist}. The median of the redshift
distribution is $<z>=0.73$. Galaxies are identified up to
$z=4.63$. As described in \cite{lefevre04} and \cite{paltani},
we have been successful in breaking into the
``redshift desert'' artificially produced by the
difficulty to identify redshifts in the range
$1.5 \leq z \leq 3$ due to our instrumental set-up, through
extensive work on galaxy templates based on
the high redshift galaxies measured in the VVDS
(\cite{paltani}).
We extensively discuss the incompleteness of 
the VVDS sample vs. redshift in \cite{lefevre04}.

The strongest peaks in the
distribution are at redshifts
$z=0.667$, and $z=0.735$. A total of 149 galaxies are measured
in the $z=0.667$ peak and 116 galaxies $z=0.735$ peak. 
These structures are extending all
across transverse $\sim16$ Mpc of this survey
($z\sim0.7$ $\Lambda$CDM with $H_{0}=70$, $\Omega_m=0.3$,
$\Omega_{Lambda}=0.7$) 
in a wall-like pattern rather than in clusters.

\subsection{Absolute Magnitude and B-I distributions vs. redshift}

\begin{figure*}
   \centering
      \includegraphics[width=11cm]{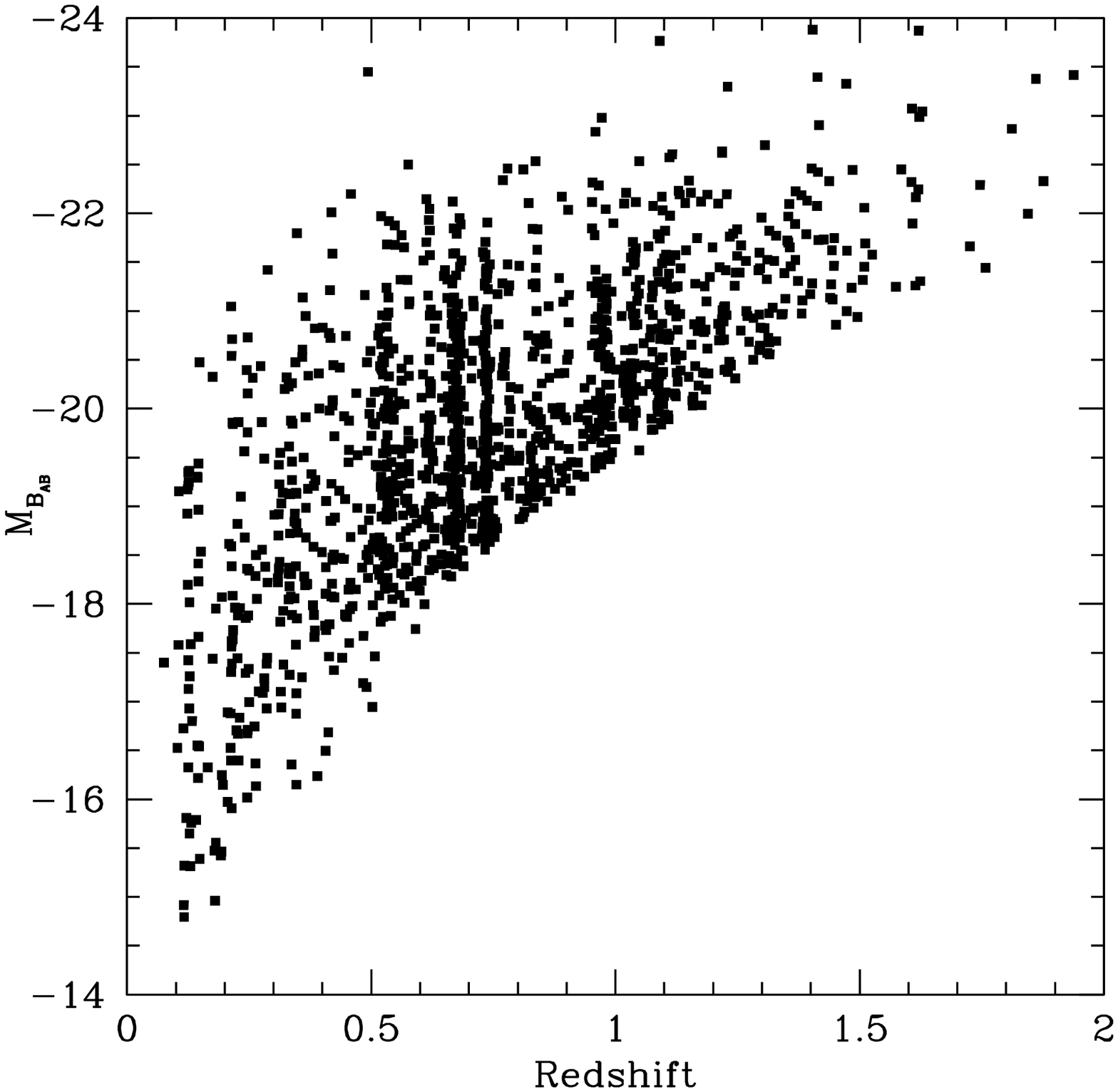}
      \caption{Absolute $M_{B_{AB}}$ magnitude - Redshift distribution 
for the full  VVDS--CDFS sample
              }
         \label{magz}
   \end{figure*}

\begin{figure*}
   \centering
      \includegraphics[width=11cm]{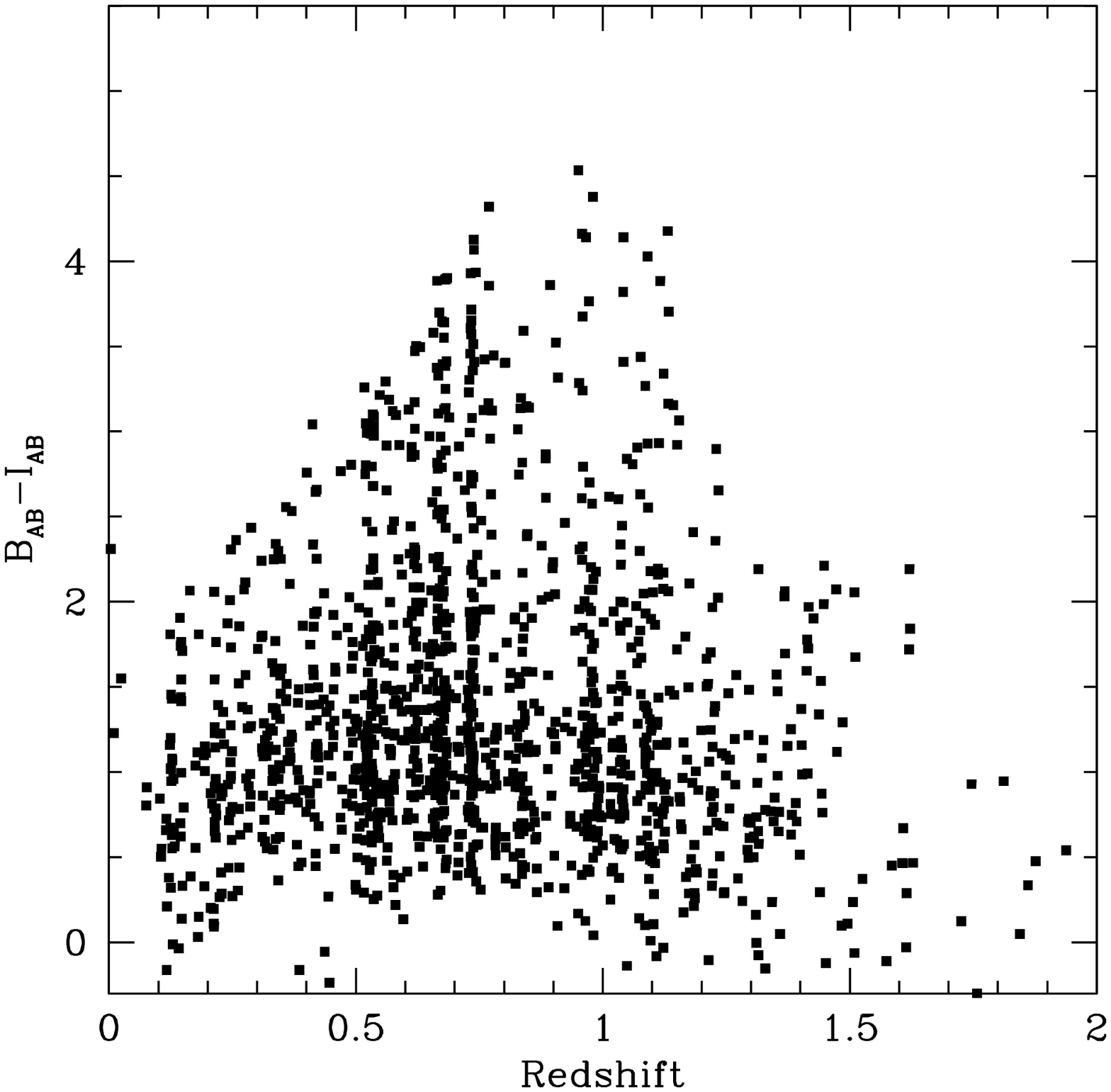}
      \caption{$B_{AB}-I_{AB}$ vs. Redshift distribution for the full 
VVDS--CDFS sample}
         \label{biz}
   \end{figure*}

The absolute magnitude $M_{B}$ vs. redshift distribution is 
presented in Figure \ref{magz}. The absolute magnitudes
have been computed based on k(z) corrections derived
from the fitting of the broad band photometry 
using rest frame galaxy templates (see \cite{ilbert}).
The $B_{AB}-I_{AB}$ vs. redshift distribution is presented 
in Figure \ref{biz}

\section{The CDFS catalog}

Our catalog contains 
1599 spectra including 1452 galaxies, 139 stars, and 8 QSOs,
with 141 of these having an uncertain
redshift identification.
We  have listed for each observed object:\\
-- the ESO Imaging Survey identification number,
equatorial coordinates $\alpha_{2000}$, $\delta_{2000}$, 
$I_{AB}$ magnitudes and associated errors (\cite{EIS})\\
-- the HST-GOODS identification number and
BVIz magnitudes and associated errors (\cite{goods})\\
-- the redshift measured by the VVDS team\\
-- the redshift quality flag (see \cite{lefevre04} 
 for a detailed description).
The flags definition from the
Canada France Redshift Survey (\cite{lefevre95})
is as follows:\\
~~~~~~~~~0: no redshift could be measured\\
~~~~~~~~~1: some spectral features like e.g. weak line(s)
or continuum break give
a possible indication of the redshift (more than $\sim50$\% 
confidence in the measurement)\\
~~~~~~~~~2: a few secure features are identified in support 
of the redshift measurement (more than $\sim75$\% 
confidence in the measurement)\\
~~~~~~~~~3: many secure features 
are identified (more than $\sim95$\% 
confidence in the measurement)\\
~~~~~~~~~4: strong secure features with
high S/N are identified (100\% confidence in
the measurement)

\section{Public data release}

We are publicly releasing all redshift measurements 
through the CENCOS (CENtre de COSmologie) database environment
on our web site http://cencosw.oamp.fr with access
to the database built under the Oracle environment. The catalog
can be searched by coordinates, redshift interval, identification
number in the EIS or GOODS catalogs, in combination with
the spectra quality flags. Upon query, the database engine
returns a list of targets, each of them can be examined
in one single summary panel with all the VVDS spectroscopy
information including the spectra, as well as the EIS and B,V,I,z HST-GOODS 
images, and associated photometry as shown in Figure \ref{vdb}.

   \begin{figure*}
   \centering
      \caption{VVDS database output panel for one single object. 
All information is presented, including the EIS and HST-GOODS
identifiers, the ground based and HST-ACS magnitudes, the 
VVDS redshift and quality flag, and the HST-GOODS images
and VVDS spectrum are displayed. 
              }
         \label{vdb}
   \end{figure*}

\section{Summary}

In the framework of the VIMOS VLT Deep Survey (VVDS),
we have observed a large sample of galaxies around
the Chandra Deep Field South, and are releasing the
redshift data to the community. A total of 1599
objects with $I_{AB} \leq 24$ have a measured redshift.
The completeness in redshift measurement for the targeted
objects is high, above 88\%. We find that the 
redshift distribution has a median of $z=0.73$,
with strong high density peaks observed across the field.

The combination of this redshift survey and the 
HST-ACS GOODS survey enables detailed studies of the evolution
of galaxies in the Chandra Deep Field South. 

\begin{acknowledgements}
We are grateful to A. Cimatti and the K20 team for 
releasing their redshift measurements to us for
comparison.
The VLT-VIMOS observations have been carried out on garanteed 
observing time (GTO) allocated by the European Southern Observatory
as a compensation for the manpower invested by the 
VIRMOS consortium in the design, manufacturing and testing
of VIMOS under a contractual agreement between the 
Centre National de la Recherche Scientifique of France,
and the European Southern Observatory.
\end{acknowledgements}

\end{document}